\documentclass{aa}
\usepackage{graphics}
\begin{document}

    \thesaurus{09(09.01.1;09.16.1;10.01.1)}

\title{ ON THE ABUNDANCE GRADIENT \\OF THE GALACTIC DISK }

\author{ L. P. Martins \and S. M. M. Viegas}

\offprints{L. P. Martins}
\mail{lucimara@iagusp.usp.br}

\institute{ Instituto Astron\^omico e Geof\'\i sico, S\~ao Paulo, Brazil} 


\maketitle

\begin{abstract}
Estimates of the gas temperature in planetary nebulae obtained
from the [O III] emission line ratio and from the Balmer discontinuity
indicate differences reaching up to 6000 K (Liu and Danziger 1993).
The [O III] temperature is commonly used to obtain the ionic fractions
of highly ionized ions, particularly the O$^{++}$ and Ne$^{++}$ ions
when using the empirical method to calculate the elemental
abundances of photoionized gas from the observed emission line
intensities. However, if the gas temperature is overestimated,
the elemental abundances may be underestimated.  In particular,
it may lead to an incorrect elemental abundance gradient  for  the Galaxy,
usually used as a constraint for the chemical evolution models.
Using Monte Carlo simulations, we calculate
the systematic error  introduced 
in the abundance gradient obtained from planetary nebulae
by an overestimation of the gas temperature. The results
indicate that the abundance gradient in the  Galaxy should be steeper
than previously assumed.

.

\end{abstract}

\section{Introduction}

Since the seminal paper by Peimbert \& Costero (1969) discussing
the empirical methods to obtain the chemical abundances, planetary 
nebulae (PN) observations have been used to derive the 
chemical composition of the
 interstellar  gas  in the Galaxy (for example, Peimbert and 
Torres-Peimbert 1971), as well as in nearby galaxies 
(for example, Ford et al. 1973). The results have been used for 
different purposes including
PN classification (Peimbert 1978, 1990, Fa\'undez-Abans \& 
Maciel 1987a) and abundance gradient determination (Fa\'undez-Abans \& 
Maciel 1986, 1987b). 

The empirical method used to derive the elemental chemical
abundance from emission lines depends on the gas temperature and 
electron density  (McCall 1984). The temperature is obtained
from the observed [O III] line ratio (T$_{OIII}$), from the [N II] line ratio 
(T$_{NII}$), or from the Balmer discontinuity (T$_{Bal}$), and usually
give different values. On the one hand, the difference between T$_{OIII}$ 
and T$_{NII}$ is probably due to the fact that O$^{++}$ and N$^{+}$ 
are in different regions, respectively at the high- and low-ionization zones. 
On the other hand, a T$_{OIII}$ higher than T$_{Bal}$ is generally explained
by the presence of temperature  fluctuations (Peimbert 1967).  
Very accurate data for a large sample of PN show that
the difference between T$_{Bal}$ and T$_{OIII}$ can reach up to 6000 K (Liu
\& Danziger 1993). As discussed by the authors, such a 
value can not be reproduced by photoionization models
of un-clumped gas. On the other hand, the presence of unresolved condensations
could solve the problem (Viegas \& Clegg 1994), indicating that T$_{Bal}$ 
is probably a better indicator of the gas temperature. In this
case, the elemental abundances must be derived assuming  T$_{Bal}$ instead
of T$_{OIII}$.

 Ionic abundances derived from both  collisionally 
 excited and recombination
lines of C and O may also indicate the presence of temperature and/or
density fluctuations in planetary nebulae, as recently discussed
by Mathis, Torres-Peimbert and Peimbert (1998). Abundances derived from
recombination lines are usually higher than those from  collisionally 
excited lines.

Regarding the abundance gradient in the Galaxy, it indicates that
the abundances are higher closer to the galactic center. Since the oxygen
lines are the main coolants, the gas temperature of the PN must be
lower closer to the galactic center. In addition, the forbidden line
emissivities increase rapidly with the gas temperature reaching 
a plateau for T $\geq$ 5 x 10$^4 $K. Thus, a change in the gas
temperature from T$_{OIII}$ to T$_{Bal}$ must induce a bigger change
in the abundance of the PN closer to the center, and, consequently, 
a change of the abundance gradient of the Galaxy. 

In this paper, we quantify the systematic error in the abundance gradient 
due to an
overestimation of the gas temperature, using a Monte Carlo method.
The data sample is discussed in \S 2. The method used and the results
are presented in \S3. The conclusions appear in \S4.

\section{The PN  sample}

As proposed by Peimbert (1978), the PN of the galactic disk can be
classified in type I, II and III. However, our sample includes
only type II PN, which are probably more representative of the galactic
chemical evolution. In fact, they are relatively young, 
produced by intermediate mass stars and participate  in  the galactic
rotation (Maciel \& Dutra 1992). On the contrary,
 type I PN are probably very young and
their chemical abundance would correspond to the present interstellar
abundance (Maciel \& K\"oppen 1994), while type III PN 
have probably originated from old less massive stars and could be
 displaced from their   birthplace (Maciel \& Dutra 1992). 

Previous gradient determinations
were obtained using the abundance values provided by 
different authors. Some of them included objects for which
the T$_{OIII}$  or the electron density  could not be calculated,
so the abundances were derived assuming a given value for these
quantities. In order to estimate the systematic error in the galatic 
abundance gradient induced by an overestimation of the temperature, 
we need  a  homogeneous sample of abundance data.  Since the data 
available in the literature come
from different observations and authors, it was necessary to recalculate
the empirical abundances for all the objects in the sample from the
observed emission-line intensities.
For this, the optical line intensities necessary
to calculate the temperature from the [O III] and [N II] line ratios, 
the  density from the [S II] line ratio, as well as the ionic fractional 
of the ions present in the gas, are needed. Therefore, 
among all the type II PN  data
in the literature, only those with those line intensities available,
as well as the galactocentric distance, were selected. These criteria
reduced the sample to 43 objects listed in Table 1 with the 
corresponding references. The adopted distances come from 
Maciel \& K\"oppen (1994).

\section{ Empirical Abundances}

We are interested in analysing systematic errors in the elemental 
abundance gradient derived from PN.
 Usually the gradient is obtained from abundance data  available  
in the literature. In our case, we need  a  homogeneous sample, i.e., 
the elemental abundance must be derived
by the same method, in particular using the same 
equations for the ionic fractions and ionization corrections.

Following Peimbert \& Costero (1969), the empirical method used
to derive the chemical abundances is based on the observed optical lines
and depends on the temperature and electron density of the emitting
region. Here the emission line used are: [O II]$\lambda$ 3727, [O II]$\lambda$ 
7327,
 [O III]$\lambda$ 4363, [O III]$\lambda$ 4959+5007, [N II]$\lambda$ 6548+6584, 
[Ne III]$\lambda$ 3868 + 3967, [S II]$\lambda$ 6717, [S II]$\lambda$ 6730, 
[S III]$\lambda$ 6312, 
He I $\lambda$ 5876, He II  $\lambda$ 4686 and H$\beta$.
It is usually assumed that the temperature of the high 
and low ionization regions are given, respectively, by the [O III]
and [N II] line ratios. Because the dispersion of most of the observations
is not enough to separate the [O II] doublet, the electron density is
obtained from the [S II] line ratio. Once the physical conditions of the
emitting regions are obtained, the ionic abundances, relative to  H$^+$,
are calculated from the observed emision-line  intensities corrected for
reddening. 

The ionic abundances for  O$^+$, O$^{++}$, N$^+$, Ne$^{++}$,
S$^+$ and S$^{++}$ have been obtained using the
emission line coefficients from McCall (1984). However, since there are
 unobserved ions present in the gas,  ionization 
correction factors are adopted in order to obtain the
elemental abundances, as shown in equations 1 to 4 below. 
The ionic abundances of He$^{+}$ and He$^{++}$ have 
been obtained from  Brocklehurst (1972) and 
the corrections for collisional de-excitation 
 of He$^{+}$ adopted from Kingdon and Ferland (1995). The total
helium abundance is the sum of the ions He$^{+}$ and He$^{++}$ since the 
neutral helium in  those objects is negligible.

\begin{equation}
 {O\over{H}} = {(O^+ + O^{++})\over {H^+} }  {\left(He\over {He^+}\right) }.
\end{equation} 
\rightline{(Peimbert \& Torres- Peimbert 1977)}

\begin{equation}
{N\over{H}} = {(N^+ )\over {H^+} }{\left(O\over {O^+}\right) }.
\end{equation}
\rightline{(Peimbert \& Torres- Peimbert 1977)}

\begin{equation}
{S\over{H}} = {(S^+ + S^{++} )\over {H^+} }{\left[1.43 + 0.196 
\left(O^{++}\over 
{O^+}\right)^{1.29}\right] }.
\end{equation}
\rightline{(K\"oppen et al. 1991)}

\begin{equation}
{Ne\over{H}} = {\left(Ne^{++} \over {H^+}\right) }{O^+ + O^{++}\over 
{O^{++}} }.
\end{equation}
\rightline{(Peimbert 1990)}

\bigskip

The calculated elemental abundances are listed in Table 1. 
These values are used to derive the
standard abundance gradient, $\alpha_0$, in the absence of temperature 
fluctuations.

\begin{table*} 
\caption[]{Results of density, temperatures and abundances}
\begin{tabular}{c  c  c  c  c  c  c  c  c  c} 
\hline
\noalign{\smallskip}
 Nebulae  & $Ref^{[\mathrm{a}]}$  &  n$_{e}$ & $T_{NII}$ & $T_{OIII}$ 
&O/H & N/H & S/H & Ne/H & $R(Kpc)^{[\mathrm{b}]}$ \\
\noalign{\smallskip}
\hline
\noalign{\smallskip}

NGC 2371&2&3114&9638&15919&8.60&8.31&7.42&7.53&9.90\\ 
NGC 2392&2&4469&7316&13676&8.82&8.63&7.73&7.80&10.33 \\
NGC 2867&8&3474&9981&11181&8.82&8.27&7.31&7.97&8.42 \\
NGC 3918&1&7320&9351&12065&8.78&8.52&7.64&7.70&7.84 \\
NGC 5882&6&4994&9897&8979&8.76&7.78&7.47&--&7.22 \\
NGC 6210&7&3557&12290&9918&8.50&7.88&7.45&7.87&7.78\\ 
NGC 6309&2&4828&10151&11267&8.84&8.35&7.92&7.80&6.51 \\
NGC 6439&6&6295&8767&8909&9.11&8.62&7.27&8.50&4.84 \\
NGC 6543&2&4329&10040&8249&8.70&8.05&7.54&8.99&8.59 \\
NGC 6563&6&5210&11174&12032&8.43&8.35&--&--&6.62 \\
NGC 6565&4&2246&9710&10383&8.84&8.52&7.24&8.20&7.01\\ 
NGC 6572&2&12076&6074&9802&8.91&8.10&7.55&8.20&7.87 \\
NGC 6578&3&5438&11059&8371&8.76&7.95&--&8.21&6.45 \\
NGC 6720&7&825&9780&11120&8.69&8.49&7.25&8.05&8.22 \\
NGC 6790&2&13468&19110&11683&8.58&7.93&7.32&7.76&7.38\\ 
NGC 6818&2&1742&11677&12841&8.95&8.54&7.67&7.66&7.24 \\
NGC 6826&2&2903&12594&10569&8.37&7.28&6.81&6.89&8.45 \\
NGC 6879&4&6967&14640&10129&8.62&7.75&7.47&7.96&7.21 \\
NGC 6884&2&7928&12507&10859&8.71&7.99&7.27&7.90&8.44 \\
NGC 6886&2&13446&11022&11850&8.86&8.31&6.95&8.02&7.80 \\
NGC 6894&3&383&14815&8219&8.79&8.53&7.39&8.29&8.10 \\
NGC 7026&2&11664&9990&9003&8.80&8.46&7.38&8.29&8.53 \\
NGC 7662&2&3623&10113&13591&8.62&8.00&7.80&7.49&8.75 \\
IC 418&6&13058&8456&13121&8.60&7.83&6.53&--&9.73 \\
IC 1297&4&3478&8924&10098&8.88&8.73&7.98&8.07&5.71 \\
IC 2003&2&8517&16316&11593&8.64&8.13&7.19&7.65&10.72 \\
IC 2149&5&4754&9103&9727&8.93&7.10&--&8.11&9.55 \\
IC 2165&2&5587&12023&14067&8.61&8.03&6.93&7.48&9.97\\ 
IC 2501&6&40972&9451&9516&8.73&8.16&6.89&--&8.38 \\
IC 2621&6&18979&12428&10994&8.98&8.67&7.26&--&7.97 \\
IC 4776&2&14651&15865&8564&8.79&8.02&7.53&8.03&5.29 \\
IC 5217&2&12965&12186&11230&8.57&8.00&7.35&7.85&9.42 \\
He2-37&8&270&10169&12824&9.05&8.59&7.13&8.01&8.62 \\
He 2-48&8&196&11235&11820&8.57&8.05&7.00&7.89&8.87 \\
He 2-115&6&21032&12650&12384&8.13&7.52&6.30&--&7.05 \\
He 2-141&6&2916&10761&15015&8.78&8.31&6.89&--&6.40 \\
Hu 1-1&2&2012&10278&12883&8.62&8.06&6.95&7.93&11.55 \\
J 320&2&4816&12158&12456&8.39&7.66&7.29&7.75&12.36 \\
J 900&2&4521&11054&12167&8.65&8.02&6.88&7.69&10.55 \\
M 1-4&4&6975&11002&12077&8.43&7.61&7.25&7.76&9.97 \\
M 1-5&5&2121&12251&15493&7.96&7.40&6.32&--&10.59 \\
M 1-54&5&2074&9023&9541&8.90&8.69&7.34&--&5.51 \\
Th 2-a&8&1466&12435&11840&8.89&8.50&--&8.02&7.30 \\

\noalign{\smallskip}
\hline
\noalign{\smallskip}

\end{tabular}
\begin{list}{}{}
\item[${\mathrm{a}}$]  References: (1)  Torres-Peimbert \& Peimbert 1977; 
(2) Aller \& Czyzak 1983;
(3) Aller \& Keyes 1987; (4)  Kaller et al. 1997;(5) Barker 1978;
(6) Freitas-Pacheco et al. 1992; (7) French 1981; (8) Kingsburgh \& Barlow 
1992.
\item[${\mathrm{b}}$]  Distances: Maciel \& K\"oppen 1984
\end{list} 
\end{table*} 
\begin{figure*}
\resizebox{\hsize}{!}{\includegraphics{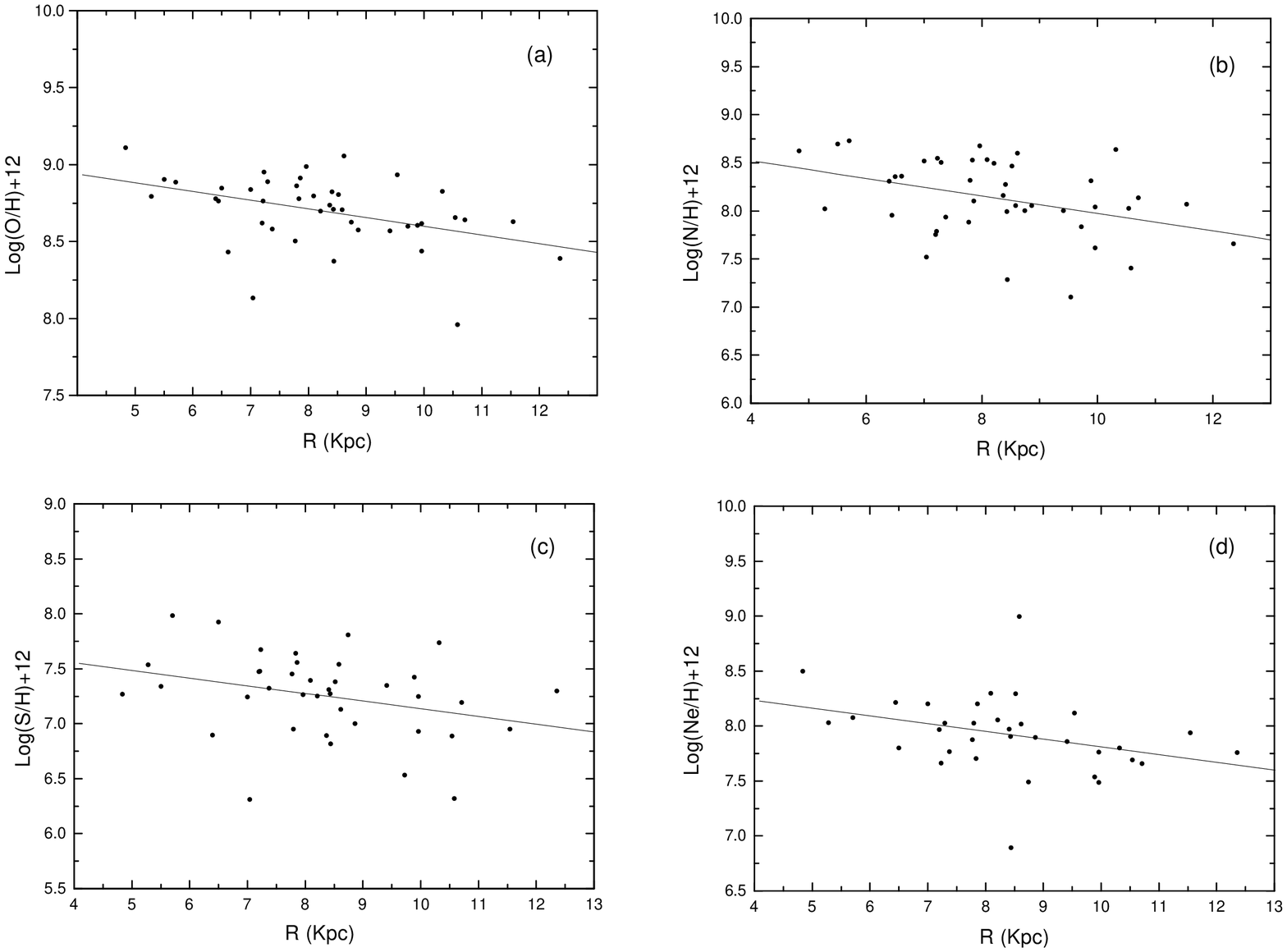}}
\caption{Radial abundance gradients: (a) O/H, (b) N/H, (c) S/H and (d) Ne/H. The 
solid line
corresponds to the linear fit to the data. }
\label{grad}
\end{figure*}

Notice that some values 
may differ from those given in the literature. The reason for the 
different results are mainly due to the collisional term included 
in the estimate of the He$^+$ fractional 
abundance and in the icf value used for S/H. 

The He$^+$ fractional abundance is
used as the icf correction for  the O abundance (Eq. 1), 
and an incorrect value may afffect all the results derived from it. 
The main problem comes from collisional correction of He$^+$.
In some of the  previous papers this correction is not accounted 
for (Freitas Pacheco et al. 1992), leading to an overabundance of 
the He$^+$ fractional abundance, and consequently,
of the He abundance. Other authors accounted for the collisonal correction 
(K\"oppen et al. 1991) as proposed by Clegg (1987) which may overcorrect 
He$^+$/He (Kingdon and Ferland 1995 ), leading
to an underestimate of the He abundance. However, for the objects
listed in Table 1, the results using Clegg's correction or Kingdom 
and Ferland correction for He I $\lambda$5876 differ by less than 3 \%.

Regarding the S abundance, the icf used by different authors 
has changed over  the years leading to different results. 
Two decades ago,  
an icf similar to the icf   for N
was assumed  (Barker 1978), followed by a more precise value ( Barker  1983).
However, several authors (for example Freitas Pacheco et al. 1991 
and references therein) used the icf value suggested by Dennefeld and Stasinska
(1983), based on photoionization models for HII regions. Because the 
HII region ionizing stars have lower temperatures, the high 
ionization zone of HII regions is smaller than
in planetary nebulae, leading to a smaller ionization correction factor due to 
the presence of highly ionized ions. Thus, when the icf derived for 
HII regions  is applied to PN  it systematically gives lower  
S abundances than those obtained using the icf proposed by K\"oppen et 
al. (1991), obtained from an extensive
grid of density bounded photoionization models for planetary nebulae.

\section{Abundance gradients}
For each element (O, N, Ne and S), the radial gradient is obtained 
from a linear fit of the elemental abundance versus distance 
(Figure 1a,b,c and d). The results are listed in Table 2.

\begin{table} 
\caption[]{Coeficients of the linear $fits^{[{\mathrm{a}}]}$}
\begin{tabular}{c c c c c} 
\hline
\noalign{\smallskip}
      & O    & N   & S   & Ne\\
\noalign{\smallskip}
\hline
\noalign{\smallskip}

$\alpha_{0}$  & -0.054   &-0.084   & -0.064   & -0.069  \\
$\sigma$($\alpha$$_{0}$) & 0.018& 0.084&0.035&0.034\\
$\beta$ & 9.16&8.86&7.83&8.51\\
$\sigma$($\beta)$ & 0.16&0.29&0.30&0.30\\
r&-0.42&-0.39&-0.31&-0.34\\
N&43&43&39&34\\

\noalign{\smallskip}
\hline
\noalign{\smallskip}
\end{tabular}
\begin{list}{}{}
\item [${\mathrm{a}}$]  Log(X/H)+12 =$\alpha$$_{0}$R + $\beta$; r is the 
correlation coeficient and N is the 
number of data points used  .  
\end{list}

\end{table}


\begin{table*}[t]
\caption[]{Comparation between the gradients of this paper and  those found 
in the $literature ^{[{\mathrm{a}}]}$}
\begin{tabular}{cccccc} 
\hline
\noalign{\smallskip}
    & This paper   & MK94    & FM86   &PP93   & MQ99\\
\noalign{\smallskip}
\hline
\noalign{\smallskip}

$\alpha_{O}$ & -0.054 $\pm$ 0.019 &-0.069 $\pm$ 0.006 & -0.072 $\pm$ 0.012 & 
-0.03 
$\pm$ 0.01 & -0.058 $\pm$ 0.007\\
$\alpha_{N}$ & -0.084 $\pm$ 0.034 &-- & -0.072 $\pm$ 0.028 & -0.05 $\pm$ 0.01 
&-- \\
$\alpha_{S}$ & -0.064 $\pm$ 0.035 &-0.067 $\pm$ 0.006 & -0.098 $\pm$ 0.022 & -- 
& -0.077 $\pm$ 0.011 \\
$\alpha_{Ne}$ & -0.069 $\pm$ 0.034 &-0.056 $\pm$ 0.007 &--  & -0.05 $\pm$ 0.02 & 
-0.036 $\pm$ 0.010 \\

\noalign{\smallskip}
\hline
\noalign{\smallskip}
\end{tabular}

\begin{list}{}{}
\item[${\mathrm{a}}$](MK94){ Maciel \& K\"oppen 1994; (FM86) 
Fa\`undez-Abans \& Maciel 1986; (PP93) 
Pasquali \& Perinoto 1993;(MQ99) Maciel \& Quireza 1999}
\end{list}
\end{table*}

The values obtained for the elemental abundance gradients are compared 
to those from previous works in Table 3. Our results have a 
larger statistical error because of the smaller number of objects 
used in this paper. Notice, however, that the results obtained 
by other authors come from a non-homogeneous
sample of elemental abundance data, where collisional correction 
for He may or may not be included and the icf for S may differ 
from one object to another. Thus  a   small
statistical error due to a larger number of objects included 
in their sample may be  misleading and hide a larger uncertainty.

In the case of neon, the icf is usually the same in all works. 
However our 
value for the gradient is barely in agreement with  the  Maciel
and Quireza (1999) result.  The PN sample used by these 
authors include 4 PN with distance from the galactic center   larger
than 12 kpc, whereas all the PN in our sample are closer than 
12 kpc. Three of these  distant  PN are usually classified as type I planetary.
However, they were reclassified as type II by Maciel and Quireza (1999)
and included in their sample. Since they have high Ne abundance, their Ne
abundance gradient is flatter. Without these PN in the sample, the
Ne gradient is -0.042 $\pm$ 0.014 (Quireza 1999), which is in agreement with our 
result (Table 3) within the errors.

\subsection{ Effect of the gas temperature}

If the Balmer temperature were available for most of the type II PN,
a new value for the galactic abundance gradient could easily be
obtained by recalculating the chemical abundances for each object
assuming T$_{Bal}$ as the gas temperature. As shown by Viegas \& Clegg
(1994), if the difference between T$_{OIII}$ and T$_{Bal}$  is due to density
fluctuations, the oxygen and neon abundance may increase up to 50 \%. This
would  resolve  the discrepancy between the elemental abundances 
derived from  permitted lines and from forbidden lines.

The value of  T$_{Bal}$ is not available for most of the PN of our sample, thus
the estimation of the systematic error, introduced into the abundance
gradient by an uncertainty in the gas temperature, is obtained by
Monte Carlo simulations. The method is similar to that used by Steigman,
 Viegas and Gruenwald (1997). For each PN, we assume that the T$_{OIII}$ 
is  overestimated
by $\Delta$T chosen from a distribution ranging from 
zero to $\Delta$ T$_{max}$,
following a probability P($\Delta$T), which can be constant,
 linear increasing or linear decreasing. Thus, if a constant  P($\Delta$T) is
assumed for each object,  any value of $\Delta$T between 0 and
$\Delta$T$_{max}$  has the same probability to be randomly chosen. On the
other hand, if a linear increasing (or deacreasing) probablility is
assumed, higher (or lower) $\Delta$T values are favored.

Once the type of probability and 
$\Delta$ T$_{max}$ are chosen, the chemical 
abundances are recalculated for each PN in the sample
using T = T$_{OIII}$ - $\Delta$T for the high ionization zone,
as described in \S 2.1. A  new value of the elemental abundance gradient,
 $\alpha$, is then obtained by linear fit for each element,
as well as the difference $\Delta$$\alpha$=$\alpha$ - $\alpha_0$.  
The procedure 
is repeated 15,000 times and the $\Delta$$\alpha$ average value 
gives the estimate of the systematic error in the gradient due
to an overestimation of the gas temperature.

Since the  difference between T$_{OIII}$ and T$_{Bal}$ 
is not easily explained, a possible overestimation of T$_{NII}$ 
must be also analized. 
There is no reason to adopt the same change $\Delta$T for T$_{OIII}$ 
and T$_{NII}$. In fact, no correlation was found between these
two temperatures (Fig. 2). In addition, for most of the PN, T$_{NII}$ 
is close to T$_{Bal}$. Thus, when calculating the systematic error
in the abundance gradient,  only a decrease in T$_{OIII}$ is accounted for;
T$_{NII}$ remaining constant.


\begin{figure}[b]
\resizebox{\hsize}{!}{\includegraphics{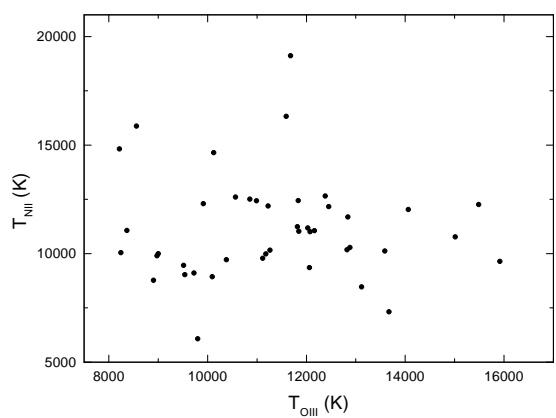}}
\caption{Plot of the temperatures $T_{NII}$ versus $T_{OIII}$. 
Figure shows that exists no
correlation between these two temperatures. The correlation coeficient is 0.21}
\end{figure}

\subsection{ Systematic error}

The PN sample observed by Liu \& Danziger (1993) shows that 
the difference between T$_{OIII}$ and T$_{Bal}$ can reach up to 6000 K,
although most objects show a difference less than 4000 K.  This value will 
be 
assumed as the maximum in our calculations.

The results of the Monte Carlo simulations are shown in Table 4 and 5,
 for $\Delta$T$_{max}$ equal to  4000 K and  2000 K, respectively. 
In both cases, the results obtained with a lower T$_{OIII}$ is to 
steepen the gradients.

Because of the rapid increase of the line emissivity with the 
gas temperature and of the expected increase of the PN gas temperature
from inner region to the outer region of the Galaxy, we expect that the
increase of the abundance, due to a  decreasing of T$_{OIII}$,
is stronger for the PN closer to the center, leading to a steeper gradient.

This effect is found for all elements. Although the N and S abundances 
are not directly dependent on 
T$_{OIII}$ (as O and Ne abundances  are), steeper gradients
are also obtained. It is a second order effect, because  
a decrease of T$_{OIII}$ induces a change in the icf of N and S.


\begin{table}[h]
\caption[]{Results of  the Monte Carlo with  $\Delta$T$_{max}$ = 4000K}
\begin{tabular}{cccc}
\hline
\noalign{\smallskip}
  & &$\Delta$$\alpha$ &   \\
\hline
\noalign{\smallskip}
  & P constant   &  P crescent    &  P decrescent\\
\noalign{\smallskip}
\hline
\noalign{\smallskip}

O & 0.039 $\pm$ 0.021 & 0.056 $\pm$ 0.018 & 0.023 $\pm$ 0.016 \\
N & 0.036 $\pm$ 0.016 & 0.051 $\pm$ 0.014 & 0.022 $\pm$ 0.012 \\
S & 0.038 $\pm$ 0.016 & 0.055 $\pm$ 0.018 & 0.022 $\pm$ 0.015 \\
Ne & 0.052 $\pm$ 0.031 & 0.074 $\pm$ 0.028 & 0.030 $\pm$ 0.023\\

\noalign{\smallskip}
\hline
\noalign{\smallskip}

\end{tabular} 

\end{table}

\begin{table}[h] 
\caption[]{Results of   the Monte Carlo with   $\Delta$T$_{max}$ = 2000K}
\begin{tabular}{cccc}
\hline
\noalign{\smallskip}
& &$\Delta$$\alpha$ &  \\
\hline
\noalign{\smallskip}
     &  P constant   &  P crescent    &  P decrescent\\
\noalign{\smallskip}
\hline
\noalign{\smallskip}
 
O & 0.015 $\pm$ 0.008 & 0.021 $\pm$ 0.007 & 0.009 $\pm$ 0.007 \\
N & 0.014 $\pm$ 0.007 & 0.020 $\pm$ 0.006 & 0.009 $\pm$ 0.005 \\
S & 0.013 $\pm$ 0.008 & 0.020 $\pm$ 0.007 & 0.008 $\pm$ 0.007 \\
Ne & 0.019 $\pm$ 0.012 & 0.026 $\pm$ 0.010 & 0.013 $\pm$ 0.010\\

\noalign{\smallskip}
\hline
\noalign{\smallskip}
\end{tabular} 
\end{table} 


\section {Concluding remarks}

The overestimation of the temperature  in planetary nebulae, used to obtain 
the elemental abundances, may lead to a systematic uncertainty
in the radial abundance gradient  of the Galaxy. Because of the lack of
observational data necessary to obtain the Balmer temperature,
the systematic  uncertainty  was evaluated by Monte Carto simulations,
where the decrease in the gas temperature for each PN in the sample
is chosen randomly between zero and $\Delta$T$_{max}$. 
The radial gradients tend to
become steeper as long as the temperature fluctuations are taken into
account. 

Several estimations of the radial gradient  of the Galaxy are available in the 
literature, obtained from  objects other than the already discussed PN.
The galactic HII regions indicate an oxigen gradient of about -0.07 dex 
Kpc$^{-1}$
(Shaver et al. 1983), very close to that obtained from the PN data,which is 
 also found from B type stars (Smartt \& Rolleston 1997,
Gummersbach et al. 1998). More recently, a new result of about -0.04
for the O abundance gradient was obtained from 
HII regions (Deharveng et al. 1999). 
The T$_{OIII}$ temperature of these HII regions are close to the 
value obtained from radio recombination lines, indicating that temperature 
fluctuations may not be present. However, the value of the O gradient
was obtained by a linear fit with a sample which includes O abundance data
 from Shaver et al. (1983), although assuming a low weight for them.
 We calculated the non-weighted O abundance gradient for the same sample and
obtained -0.052, thus closer  to  our PN result. It is clear 
that new observations are needed to increase the number of objects
for which a more precise T$_{OIII}$ can be obtained.
  
On the other hand, for open clusters the  
Fe/H gradient was -0.095 dex Kpc$^{-1}$ (Friel 1995), but recent
results indicate a flatter gradient in agreement with the O/H gradient
from the B stars (Friel 1999). The temperature effect discussed in this 
paper
could also apply to HII regions, and we would expect that the corresponding
abundance gradient would also be steeper, approaching the former value 
obtained from open clusters, although O and Fe are produced by different
progenitors.  However, two important issues are 
{\it how to explain the observed
gradient and how constant it is during the galactic evolution}. 

A value for the radial abundance
 gradients as precise as possible is
of fundamental importance for the chemical evolution models 
of our Galaxy (e.g. Chiappini 1998) The abundance 
gradient is an important
constraint  on  the models, since it is not restricted to the solar 
vicinity as are  most of the other constraints. The temporal and 
spatial behavior of the gradient depends on 
the star formation rate
and on the gas density distribution in the disc.
Regarding chemical evolution
models, different authors adopt different prescriptions for the
input parameters, and different solutions are obtained.
 Some constraints are satisfied by 
different models, however, the abundance gradient is one of the few
 that may really determine the model. The model discussed
by Chiappini (1998) gives an O gradient of - 0.04 dex/Kpc for the inner
part of the Galaxy, which is too flat. She suggests that if radial
flows are included in the model the theoretical 
O gradient could become steeper. As shown in this paper,
the real elemental abundance gradient may be steeper than
previously assumed, and it may then imply that radial flows
must really be accounted for in future models.

\begin{acknowledgements} 
 We are indebt for discussions with 
R. B. Gruenwald, W. J. Maciel and R. Costa.
We are also thankful to an anonymous referee
whose useful comments greatly improved this paper. 
This work is partially supported by grants 
from CNPq (304077/77-1), from FAPESP (98/14613-2), 
and from PRONEX/FINEP (41.96.0908.00)
\end{acknowledgements}

\end{document}